\documentclass[twocolumn]{aastex62}

\usepackage{savesym}
\savesymbol{tablenum}
\usepackage[separate-uncertainty=true,multi-part-units=single]{siunitx}
\DeclareSIUnit\year{yr}
\restoresymbol{SIX}{tablenum}

\graphicspath{{./}{figures/}}

\received{January 1, 2018}
\revised{January 7, 2018}
\accepted{\today}
\submitjournal{ApJ}

\shorttitle{Explosion Chemistry in TMC-1}
\shortauthors{Holdship et al.}

\begin{document}

\title{Investigating the Efficiency of Explosion Chemistry As A Source of Complex Organic Molecules in TMC-1}

\correspondingauthor{Jonathan Holdship}
\email{jrh@star.ucl.ac.uk}

\author[0000-0002-0786-7307]{Jonathan Holdship}
\affil{Department of Physics \& Astronomy, University College London, Gower Street, London, WC1E 6BT}

\author[0000-0002-3237-730X]{Jonathan Rawlings}
\affil{Department of Physics \& Astronomy, University College London, Gower Street, London, WC1E 6BT}

\author[0000-0001-8504-8844]{Serena Viti}
\affil{Department of Physics \& Astronomy, University College London, Gower Street, London, WC1E 6BT}

\author[0000-0001-5121-5683]{Nadia Balucani}
\affil{Dipartimento di Chimica, Universita` di Perugia, Via Elce di Sotto, 8 I-06123 Perugia, Italy}

\author{Dimitrios Skouteris}
\affil{Scuola Normale Superiore, Piazza dei Cavalieri 7, I-56126 Pisa, Italy}

\author{David Williams}
\affil{Department of Physics \& Astronomy, University College London, Gower Street, London, WC1E 6BT}



\begin{abstract}
Many species of complex organic molecules (COMs) have been observed in several astrophysical environments but it is not clear how they are produced, particularly in cold, quiescent regions. One process that has been proposed as a means to enhance the chemical complexity of the gas phase in such regions is the explosion of the ice mantles of dust grains. In this process, a build up of chemical energy in the ice is released, sublimating the ices and producing a short lived phase of high density, high temperature gas. The gas-grain chemical code UCLCHEM has been modified to treat these explosions in order to model the observed abundances of COMs towards the TMC-1 region. It is found that, based on our current understanding of the explosion mechanism and chemical pathways, the inclusion of explosions in chemical models is not warranted at this time. Explosions are not shown to improve the model's match to the observed abundances of simple species in TMC-1. Further, neither the inclusion of surface diffusion chemistry, nor explosions, results in the production of COMs with observationally inferred abundances.
\end{abstract}


\section{Introduction} \label{sec:intro}
Complex Organic Molecules (COMs) are organic molecules with six or more atoms, over 50 species of which have been detected in the ISM \citep{Herbst2009}. Understanding the chemistry that leads to the formation of such large molecules is an active area of research including laboratory experiments \citep{Chuang2016,Bergantini2017}, observational surveys \citep{Ceccarelli2017,Belloche2016}, and modelling work \citep[eg.][]{Coutens2018}. However, the major formation routes of COMs in star forming regions remains an open question.\par
It is possible that COMs form in the gas phase of star forming regions. For example, models have shown that proton transfer reactions between common ice mantle species that sublimate in hot cores can efficiently produce COMs \citep{Taquet2016}. Further, chemical models using gas phase reactions to form glycolaldehyde (HCOCH$_2$OH) can match the abundances observed in hot corinos \citep{Skouteris2018}. Recent observations of formamide towards the L1157-B1 shocked region were also well fit by shock models in which the parent species were released into the gas phase by the shock passage and then reacted in the warm, dense post-shock gas \citep{Codella2017}. Similarly, \citet{Kahane2013} found that observed formamide abundances in the protostar IRAS 16293-2422 could be reproduced using a model that assumed neutral parent species were able to react in the warm gas.\par
Alternatively, COMs in the gas phase may be best explained by grain surface formation followed by desorption into the gas phase. In this case, the grain surface acts to improve the efficiency of formation, bringing reactants together into one location and potentially lowering the energy required. Models of both a prestellar core \citep[L1544][]{Vasyunin2017,Quenard2018} and a hot corino \citep[IRAS 16293 B][]{Quenard2018} have had success implementing the diffusion-reaction mechanism of \citet{Hasegawa1992}. However, both works rely on chemical desorption \citep{Minissale2016} to release COMs into the gas phase, the efficiency of which is not well constrained.\par
Regardless of the formation path, the problem of releasing material into the gas phase remains. Gas phase formation routes require parent species to be released from the grains and surface formation requires the release of the products. In warm regions such as hot cores or shocked zones, this poses no issue. However, in cold dark clouds, it is less obvious how efficiently material can be released into the grains. In this work, the explosions of ice mantles are considered as a possible way to both enrich the gas phase with grain surface material and to open new chemical pathways.\par
It has been proposed that the ice mantles of dust grains may undergo  explosions caused by the build up and subsequent reaction of radicals in the ice \citep{Greenberg1976}. This would release stored chemical energy and could raise the temperature of the whole dust grain. If this temperature excursion is sufficiently high, the ices will sublimate explosively. To raise a dust grain to \SI{1000}{\kelvin} would require approximately \SI{12}{\kilo\joule\per\mole}, an order of magnitude less than the typical bond energy \citet{Duley2011}.\par 
An interesting consequence of these explosions is the unique chemical phase that follows. \citet{Cecchi-Pestellini2010} and \citet{Rawlings2013a} considered that in such explosions, the sublimated ice forms an expanding shell of gas which initially has the density of the pre-sublimation solid ($\sim$ \SI{e22}{\per\centi\metre\cubed}) and a temperature of \SI{1000}{\kelvin}. This phase lasts for $\sim$\SI{100}{\nano\second} as the sublimated ice expands into the wider environment but the chemical timescale is sufficiently short in such hot, dense gas that efficient three body chemistry can take place. This would lead to the formation of complex species from the released material and the chemical enrichment of the wider gas phase.\par
Whilst the possibility of these explosions forming specific molecules such as propene (CH$_2$CHCH$_3$) \citep{Rawlings2013} and methanol (CH$_3$OH) \citep{Coutens2017} have been studied, a comprehensive model of these explosions towards a dark cloud has not been produced. In this work, a gas-grain chemical model that includes explosions is used to model observations of COMs in a dark cloud with the aim of testing whether explosion chemistry is a viable route to their formation. In Section~\ref{sec:tmc-1}, the observational data is presented. In Section~\ref{sec:model}, the chemical model is described and, in Section~\ref{sec:results}, a comparison between the model and observations is presented. 
\section{TMC-1 - Observational Data}
\label{sec:tmc-1}
In order to test whether explosion chemistry is a necessary or relevant process for dark cloud chemistry, observational constraints are required. TMC-1 is a common test case for dark cloud models \citep[eg.][]{Vidal2017,Ruaud2016} and many COMs have been detected in the region  \citep{Soma2018}, making it an ideal candidate.\par
Two tests of the models are taken into consideration. First, the inclusion of explosions in the chemical model should not interfere with the gas phase chemistry of simple species. These species must be at least as well described by explosions as they are by other models. To this end, the first part of Table~\ref{table:observations} lists simple chemical species and their abundances taken from \citet{Agundez2013}. These were calculated by those authors from observed column densities using a H$_2$ column density of \SI{e22}{\per\centi\metre\squared}. \par
Second, the primary goal is to reproduce the observed abundances of COMs in TMC-1. Using the H$_2$ column density from \citet{Agundez2013}, the column density of COMs in the region have also been converted to fractional abundances. These are listed in the second part of Table~\ref{table:observations}. The column densities of methanol (CH$_3$OH), acetaldehyde (CH$_3$CHO), methyl formate (HCOOCH$_3$) and dimethyl ether (CH$_3$OCH$_3$) were taken from \citet{Soma2018}. Propene (CH$_2$CHCH$_3$) was detected by \citet{Marcelino2007}.\par
Note that \citet{Soma2015} found that the methanol emission in TMC-1 peaks in a different location to the cyanopolyyne peak. The cyanopolyyne peak is the location from which most molecular emission in the region originates but the COMs detected by \citet{Soma2018} were detected towards the methanol peak. \citet{Soma2018} argue that the detected COMs are therefore likely to form on the grain surface or from CH$_3$OH in the gas. The reason for this is that any enhancement in CH$_3$OH would naturally be accompanied by an enhancement in the other species. If explosions were responsible for forming or releasing COMs, similar behaviour would be observed. Since the physical conditions of the two peaks are broadly similar. ($n_H\sim$\SI{e4}{\per\centi\metre\cubed} and $T_k$=\SI{10}{\kelvin}) and even the methanol abundance only varies by a factor of 1.5 \citep{Soma2018}. As a result, no distinction is made between the peaks for the sake of the modelling.
\begin{table}
\centering
\caption{Species and measured abundances in TMC-1 taken from \citet{Agundez2013} unless otherwise specified.}
\begin{tabular}{cc}
\hline
Species & Fractional Abundance \\
\hline
OH & \num{3e-7}\\
CO & \num{1.7e-4}\\
HCO$^+$ & \num{9.3e-9}\\
H$_2$CO & \num{5e-8}\\
N$_2$H$^+$ & \num{2.8e-10}\\
NH$_3$ & \num{2.5e-8}\\
CS & \num{3e-9}\\
H$_2$CS & \num{7e-10}\\
OCS & \num{2.2e-9}\\
SO & \num{1.5e-9}\\

\hline
CH$_3$OH & \num{6e-9}\\
CH$_3$CHO & \num{5.5e-10}\\
HCOOCH$_3$ & \num{1.6e-10} \\
CH$_3$OCH$_3$ & \num{1.9e-10} \\
CH$_2$CHCH$_3$ & \num{4e-9} \\
\hline
\end{tabular}
\label{table:observations}
\end{table}
\section{Model}
\label{sec:model}
\subsection{The Cloud Chemistry Model}
In order to model TMC-1 and to test the effect of explosions on the chemistry of dark clouds, the gas-grain chemical code UCLCHEM\footnote{\url{uclchem.github.io}} \citep{Holdship2017} was modified. The basic dark cloud model is described in this section.\par
UCLCHEM is used to model a single point at the centre of a dark cloud. The gas starts at a hydrogen nuclei density of  \SI{e2}{\per\centi\metre\cubed} and collapses in freefall to \SI{2e4}{\per\centi\metre\cubed} at a constant temperature of \SI{10}{\kelvin}. After the collapse, the visual extinction at the cloud centre is 10 mag. Initially, the abundance of every species except for atomic elements is set to zero whilst the elemental abundances themselves are set to their solar values \citep{Asplund2009}.\par
The model follows 528 species through a network of approximately 3000 reactions. This includes species in the gas phase and in the ice mantles. Gas phase reactions from the UMIST12 database \citep{McElroy2013}, freeze out of gas phase species onto the dust grains and the non-thermal desorption of those species back into the gas phase through UV, cosmic rays and H$_2$ formation \citep{Roberts2007} are all included in the network. In addition to this, the cosmic ray induced photo-dissociation of hydrogenated species on the grain surfaces are included using efficiencies from \citet{Garrod2006}.\par
\subsection{The Explosion Model}
\label{sec:explosion-model}
The model considers the possibility that if enough chemical energy is stored in the ice mantles, it could be suddenly released and this would lead to an explosion. This is treated by considering the abundance of H atoms in the ices. If approximately 5\% of the grain material was atomic hydrogen, the energy released through H$_2$ formation would be sufficient to heat the whole grain to \SI{1000}{\kelvin} if every H atom was involved. Thus, an explosion is triggered in the model once the H abundance in the ices reaches this threshold. \par
The hydrogen required by the model is built up by assuming there is a probability ($f_H$) that when a H atom freezes out of the gas phase and onto the ices, it remains atomic rather than immediately reacting to form H$_2$ or other species. Following \citep{Rawlings2013a} and \citep{Duley2011}, a probability of 0.1 is assumed based on the retention of H atoms in amorphous carbon films found in laboratory experiments \citep{Sugai1989}.\par
The cosmic ray induced photodissociation of species in the ice mantles also contributes to the total as any abstracted H is also stored. If a portion of this abstracted H actually desorbs or the probability of H remaining atomic in the ices is less than 0.1, this model will overestimate the amount of H in the ices. In that case, the actual impact of explosions on the chemistry in TMC-1 would be overestimated by the model. \par
\begin{table}
\centering
\caption{Parameters and adopted values for the explosion model.}
\begin{tabular}{cc}
\hline
Parameter & Value \\
\hline
Initial Density & \SI{e22}{\per\centi\metre\cubed}\\
Initial Temperature ($T_0$) & \SI{e3}{\kelvin} \\
Initial Radius ($r_0$) & \SI{e-5}{\centi\metre}\\
Sound speed ($v_s$) & \SI{e4}{\centi\metre\per\second}\\
Trapping Factor ($\epsilon$) & 1.0 \\
Atomic H Retention ($f_H$) & 0.1 \\
\hline
\end{tabular}
\label{table:modelparams}
\end{table}
To model the explosion itself, the single point model is paused and the ice mantle contents are run through a separate chemical model. In this model, the pre-explosion ice mantle is considered to form an adiabatically expanding spherical shell of gas. This gas expands and gas phase chemistry occurs until the density of the cloud is reached. The material is then added to the gas phase of the main chemical model, which resumes with depleted ices.\par
The chemical network for the explosion phase consists of 143 three body reactions. Many of which involve radicals which build up in the ices through partial hydrogenation of frozen species and photodissociation of larger species. Due to the high density, it is assumed that the reactions take place in the high pressure limit, that is to say the rates are not limited by the concentration of the stabilizing third body and the reaction proceeds at the two body rate \citep[Chapter 9][]{Jacob1999}. All reactions are listed in Table~\ref{table:explosionreactions1}. Where possible the rate coefficients are taken from the literature, otherwise rates are randomly sampled in log-space from the range \num{e-15} to \SI{e-9}{\centi\metre\cubed\per\second}. The model is then run 1000 times to generate a mean abundance and variance due to the unknown rates.\par
The parameters used for the explosion phase are listed in Table~\ref{table:modelparams}. The density and temperature of the exploding material have a time dependence based on the adiabatic expansion of a spherical shell, following the work of \citet{Cecchi-Pestellini2010}. If the shell is assumed to expand at the sound speed of the gas, then by mass conservation the density is given by,
\begin{equation}
\frac{n}{n_0} = \left(\frac{r_0}{r_0+\epsilon v_st}\right)^3
\label{eq:density}
\end{equation}
where $n$ is the number density, $r$ is the radius of the shell and the subscript 0 indicates the value of a variable at the start of the explosion. $v_s$ is the sound speed and $\epsilon$ is the trapping factor: an arbitrary constant that allows the expansion to be made slower than that of a freely expanding sphere of gas. Assuming an adiabatic expansion, the temperature, $T$, is given by,
\begin{equation}
T = T_0\left(\frac{r_0}{r_0+\epsilon v_st}\right)
\label{eq:temperature}
\end{equation}
where $T_0$ is the initial temperature, taken to be \SI{1000}{\kelvin}. This value is chosen as previous work on explosions showed that dust grains heated to this temperature could provide explanations for infrared emission bands in interstellar spectra \citep{Duley2011} and the high excitation H$_2$ emission in diffuse clouds \cite{Cecchi-Pestellini2012}.\par
\begin{figure}
\includegraphics[width=0.5\textwidth]{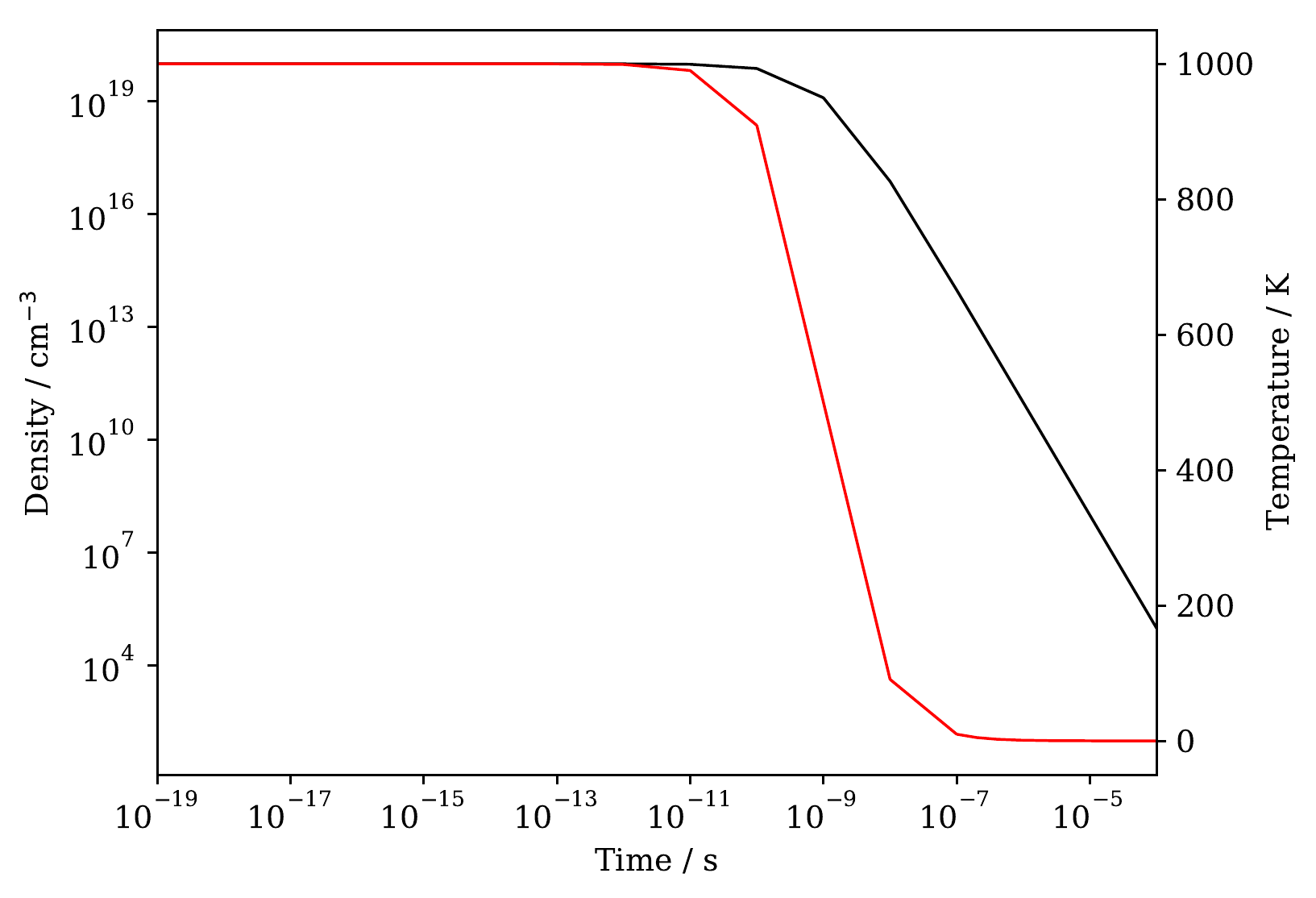}
\caption{The density (black) and temperature (red) profiles of the expanding gas shell as a function of time during an explosion.\label{fig:exp-physical}}
\end{figure}

Equations~\ref{eq:density} and~\ref{eq:temperature} are plotted in Figure~\ref{fig:exp-physical} for an $\epsilon$ of 1, the value adopted for this work. A smaller trapping factor increases the timescale of the explosion but it was found that models with $\epsilon = 0.1 $ did not produce greatly different abundances. The explosion ends when the exploding gas reaches ambient gas density. At the completion of this explosion, the abundances of the former ice mantle are added to the gas phase and the main chemical model continues.\par
\subsection{The Diffusion Model}
In order to test whether explosions are necessary to explain the abundance of COMs in TMC-1, a comparison model is employed. The explosions are turned off and the reaction of species on the grain through the Langmuir-Hinshelwood mechanism is considered. These are reactions between adsorbed molecules as they diffuse around the grain surface and they are implemented through the formalism described by \citet{Hasegawa1992}. Reaction-diffusion competition \citep[eg.][]{Chang2007} and chemical desorption \citep{Minissale2016} are also included in the model. Due to the chemical desorption, a fraction of any products created on the surface through exothermic reactions are released into the gas phase. The implementation of these processes in UCLCHEM was developed by \citet{Quenard2018}  and is extensively described in Appendix A of that work.\par
The network used for this model mainly consists of the successive hydrogenation of key species such as CO through to CH$_3$OH as well as the formation of species such as CO$_2$ from CO and O. However, the main additions to the network of \citet{Quenard2018} are reactions taken from \citet{Garrod2006} that produce the COMs in Table~\ref{table:observations}. These reactions are presented in Table~\ref{table:garrod} Each reaction is assumed to be barrierless as they are radical-radical reactions and therefore the rate is largely dependent on the diffusion rate of the reactants. 
\begin{table}
\centering
\caption{Surface Reactions necessary to produce observed COMs using the diffusion model. Reactions are taken from \citep{Garrod2006} or invented. A \# indicates a species on the surface.}
\begin{tabular}{cccc}
\hline
Reactant 1 & Reactant 2 & Product & Source\\
\hline
\#HCO & \#CH$_3$O & \#HCOOCH$_3$ & G\&H 2006\\
\#HCO & \#CH$_2$OH & \#HCOOCH$_3$ & \\
\#CH$_3$ & \#CH$_3$O & \#CH$_3$OCH$_3$ & G\&H 2006 \\
\#HCO & \#OH & \#HCOOH & G\&H 2006\\
\#CH$_3$ & \#C$_2$H$_3$ & \#CH$_3$CHCH$_2$ & \\
\#CH$_3$ & \#HCO & \#CH$_3$CHO & \\
\hline
\end{tabular}
\label{table:garrod}
\end{table}
\section{Results}
\label{sec:results}
\subsection{Effect of Explosions on Cloud Chemistry}
\label{sec:simple-results}
There are two motivating reasons to test the ability of the explosion model to reproduce observed abundances of simple species. The first is that there is the potential that the regular release of the ice mantles into the gas phase completely changes the abundances of those species. The model must reproduce the observations at least as well as a standard UCLCHEM model. Otherwise, it cannot be correct, even if it efficiently produces COMs.\par
Secondly, there are a large number of free parameters in the model, both in the assumed properties of TMC-1 and in the explosion itself. By adjusting the cloud parameters to best fit the observed abundances of simple species, the number of free parameters available to fit the COMs are reduced.\par
To fit the simple species, the so-called distance of disagreement measure \citep{Wakelam2006} was used. This is the average log difference between the model and observations. The UV flux, cosmic ray ionization rate and temperature were fit by minimizing this statistic. The temperature was varied between 0 and \SI{30}{\kelvin}. The standard cosmic ray ionization rate was taken to be \SI{1.3e-17}{\per\second} and both it and the UV flux and were varied between 
0 and 100 times the standard values. The parameter space was also sampled, repeating parameter values in proportion to the value of the distance of disagreement they produced to test the sensitivity of the abundances to the parameters.\par
\begin{figure*}
\includegraphics[width=\textwidth]{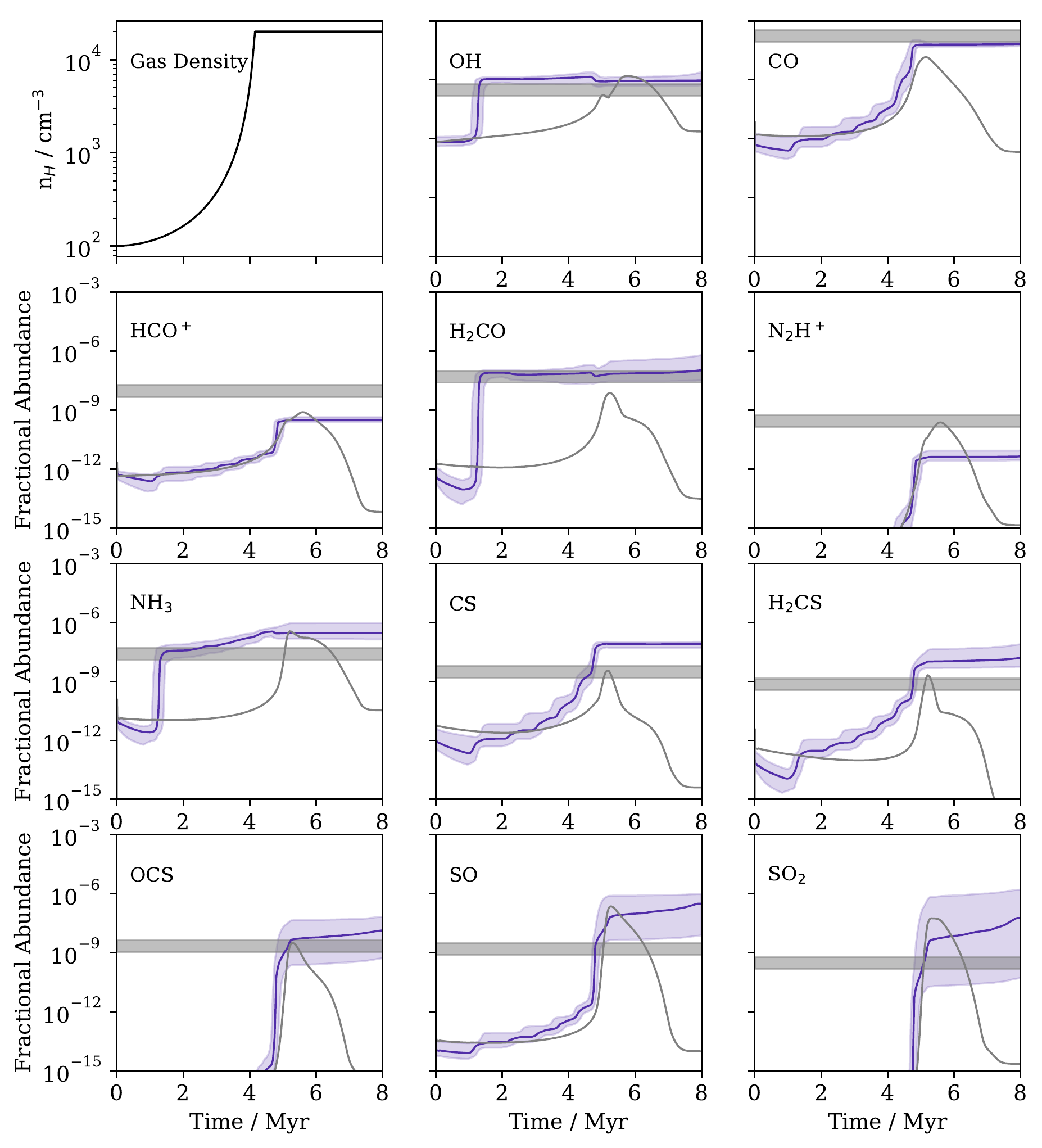}
\caption{The abundances of several simple species as a function of time in the explosion model. The purple shaded region shows the 67\% confidence interval of the abundances considering the uncertainty in the fitting. The average is plotted as a purple line and the output of a standard dark cloud UCLCHEM model without explosions is plotted in grey. The grey horizontal band in each case is the observed abundance in TMC-1 with a 0.3 dex uncertainty.\label{fig:exp-simple}}
\end{figure*}
Figure~\ref{fig:exp-simple} shows the observed abundances with the 0.3 dex uncertainty assumed by \citet{Agundez2013} in grey and the abundances obtained by the model in purple. In each subplot, the purple line shows the median abundance from the model sampling and the shaded region is given by the difference between the 17th and 83rd percentile values of the abundances across the models. The best fit is a cosmic ray ionization rate of \SI{1.7e-17}{\per\second}, a temperature of \SI{12.1}{\kelvin}, and a UV radiation field of 0.7 Habing. The parameter ranges corresponding to the shaded regions include gas temperatures between 9 and \SI{21}{\kelvin}, UV fields between 0.5 and 4.1 Habing and cosmic ray ionization rates up to 8 times the standard.\par 
Figure~\ref{fig:exp-simple} also shows the abundance of each species as a function of time in a standard dark cloud model without explosions. The major difference is that for many species, once a maximum value is reached at high densities in the standard model, freeze out starts to deplete its abundance. In the explosion model, a quasi steady state is instead reached, with explosions regularly releasing material back into the gas phase. \par
There is a problem with the model in that it does not well reproduce the observed abundances of ions. It is not uncommon for single point models of dark clouds to give low abundances of ions as they do not capture the chemistry of regions with lower visual extinction. For example, the model without explosions has a HCO$^+$ peak that is an order of magnitude too low but is within a factor of a few of that found in other dark cloud models \citep{Iqbal2018}. However, the explosions seem to exacerbate the issue and the explosion cycle averaged abundance is much lower than the non-explosion peak, particularly in the case of N$_2$H$^+$. However, given the generally good agreement between the explosion model and the observations, the model is considered to give a good representation of dark cloud chemistry.\par
It should be noted that the fact that the explosions affect the abundances of all species, even those mostly formed in the gas phase, poses a problem for the model. As noted in Section~\ref{sec:tmc-1}, observations show that different species peak in emission at different positions in TMC-1. The usual explanation is that differences between gas-phase and surface chemistry are the cause. Explosions do not present a solution to this since all species are affected similarly.\par
\subsection{COMs}
\begin{figure*}
\includegraphics[width=0.9\textwidth]{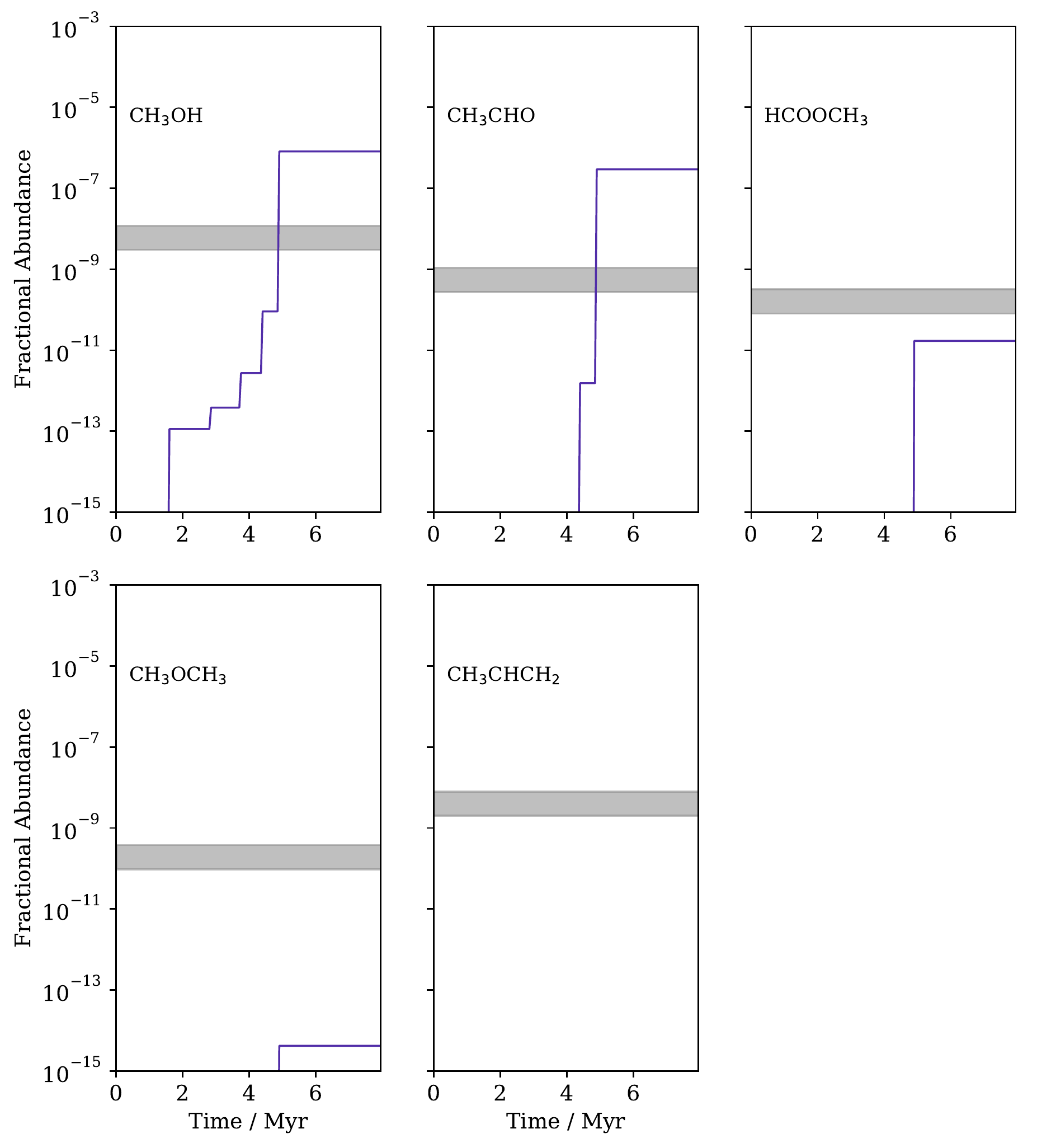}
\caption{The abundances of COMs observed in TMC-1. The horizontal bands show observed values, the purple line shows the explosion model abundances. There is a shaded region showing the results from 1000 models using random rates for the explosion reactions with unknown rates. However, they do not affect the abundance of the species shown here and so the region is not visible. \label{fig:exp-coms}}
\end{figure*}

The aim of introducing explosions into this model was to reproduce the abundance of COMs in TMC-1. In this section, the model is further compared to the abundances in the lower half of Table~\ref{table:observations}. In Figure~\ref{fig:exp-coms}, the abundances of those COMs obtained through this modelling are plotted along with the observed values. In this plot, the purple line gives the average abundance of each species in the models, having run the model many times to randomly sample unknown rates. The shaded region is not visible in the plot due to the fact that the abundances of the displayed species are unaffected by the unknown rates. \par
As can be seen in the Figure, CH$_3$OCH$_3$ and CH$_3$CHCH$_2$ are not efficiently produced in the model. The low production of these species illustrates an overall problem with the explosion model which is the short timescale of the explosion event. Unless the rate of a reaction is very high, the overall change in reactant abundances is low. In general, the proportion of an ice phase species that reacts during an explosion event is $\ll$1\%. For example, 99.99\% of HCO in the ice phase is released into the gas phase after a typical explosion and only 0.01\% reacts to form other species. Thus, the limiting factor in the formation of a COM such as CH$_3$OCH$_3$ is the rate of reaction in the explosion, not the availability of parent species.\par
This low rate of production is exacerbated by destruction in the gas phase. For example, in the reference model, CH$_3$OCH$_3$ can have a fractional abundance $\sim$\num{e-11} immediately after an explosion. If such abundances were preserved between explosions, the cumulative abundance could reach observed values. However, CH$_3$OCH$_3$ is efficiently destroyed by ions in the gas phase and so does not accumulate.\par
In the model, HCOOCH$_3$ is efficiently produced and is within an order of magnitude of the observed abundance. However, the reaction to produce HCOOCH$_3$ is unconstrained in the model and so the rate is randomly sampled. Despite this, the abundance of HCOOCH$_3$ does not vary. Tests where the reaction is removed from the explosion network show that HCOOCH$_3$ is actually produced in the gas phase. The explosions contribute by releasing  parent species from the ice mantle and the reactions during the explosion are not actually directly producing HCOOCH$_3$. Given that the timescale of the explosion appears to be too short in comparison to the chemical timescales of the explosion network, this may be the main way explosions contribute to interstellar chemistry, if they do in fact contribute.\par
Finally, both CH$_3$OH and CH$_3$CHO are each at least an order of magnitude above their observed values.  This is a result of the fact that the parent species of each molecule are extremely abundant in the ices and so even with low reaction rates a large amount of each is produced. Further, a large proportion of these species is frozen onto the dust grains and the explosions release this, greatly enhancing their gas phase abundance.\par 
In summary, whilst the explosion model gives an adequate description of the dark cloud chemistry of simple species, it does not reproduce the observed abundances of this sample of COMs. The main flaw is that the reactions which form COMs are not sufficiently fast to form large amounts of the complex species in the relatively short explosions. However, a further problem is posed by the fact that for the most simple COMs that were modelled, the predicted abundances that are too high due to the release of large amounts of ice mantle material.\par
\subsection{Comparison to the Diffusion Model}
Species that freeze onto the ices are likely to diffuse and potentially react. If these processes alone are sufficient to model the abundance of COMs in TMC-1, it is questionable whether the explosion process needs to be introduced. However, if diffusion reactions are insufficient, it is possible explosions are an important process in molecular clouds. In this section the ability of the diffusion model to reproduce COMs in TMC-1 is evaluated using the standard parameters from Section~\ref{sec:simple-results}. The abundances of observed COMs in TMC-1 and the abundances obtained in the explosion and diffusion models are summarized in Table~\ref{table:abunds}.\par
The model is successful in reproducing the abundance of CH$_3$OH. For $\sim$\SI{1}{\mega\year} after the collapse to the density of the cloud, the abundance of CH$_3$OH is within an order of magnitude of the observed value. However, the CH$_3$CHO abundance is too high as it has an abundance similar to CH$_3$OH.\par
Beyond this, the diffusion model does not reproduce the observations. The abundances of HCOOCH$_3$, CH$_3$OCH$_3$ and CH$_3$CHCH$_2$ are too low by many orders of magnitude. Given that the production of the reactants that form these species is the same in both models, this must be due to the efficiency of the diffusion of these reactants. The explosion provides a means for the reactants in Table~\ref{table:garrod} to meet and react whereas most are too heavy to quickly diffuse around the grain surface, especially competing with more mobile species such as H.\par
The diffusion model is improved if temperatures of \SI{30}{\kelvin} are used. Non-negligible amounts of the three largest COMs are produced. However, HCOOCH$_3$ and CH$_3$CHCH$_2$ are still too low by over three orders of magnitude. On the other hand, CH$_3$OCH$_3$ is actually higher in these models than the observations. Thus it is possible that if the dust temperature is $\sim$\SI{30}{\kelvin}, diffusion reactions may produce COMs. However, unless the diffusion network is significantly changed, the observations towards TMC-1 still cannot be properly explained by diffusion chemistry alone.
\begin{table*}
\centering
\caption{Abundances of COMs from observations and best fit parameters of the explosion and diffusion models.}
\begin{tabular}{lccc}
\hline
Species & Observed Abundance & Explosion Model & Diffusion Model\\
\hline
CH$_3$OH & \num{6e-9} & \num{8.1e-7}& \num{4.6e-9} \\
CH$_3$CHO & \num{5.5e-10} & \num{2.9e-7}& \num{2.2e-7}\\
HCOOCH$_3$ & \num{1.6e-10} & \num{1.7e-11}& \num{3.2e-15} \\
CH$_3$OCH$_3$ & \num{1.9e-10} & \num{4.2e-15}& \num{1.1e-15} \\
CH$_2$CHCH$_3$ & \num{4e-9} & \num{2.4e-16} & \num{4.6e-22} \\
\hline
\end{tabular}
\label{table:abunds}
\end{table*}
\section{Conclusion}
Explosions of the dust grain ice mantles through the build up of radicals in the ice were added to UCLCHEM, creating a self-consistent gas-grain chemical model with explosions. These explosions cause short lived (\SI{100}{\nano\second}) phases of high density, high temperature gas in which three body reactions can occur. The ability of the model to reproduce observations of a dark molecular cloud was evaluated, with a particular focus on complex organic molecules.\par
It was found that, despite the regular enrichment of the gas phase with ice mantle species, many simple species observed in TMC-1 were well described by the model. The majority of species had model abundances within an order of magnitude of the observed abundances and the exceptions were molecular ions which are also challenging to reproduce in models without explosions. It was also possible to conclude that explosions become more significant when the cosmic ray ionization rate is increased.\par
However, the explosion model could not reproduce the observed abundances of COMs. The abundances of those that formed efficiently on the dust grains were far larger than observed due to the regular release of the ice mantles into the gas phase. Two destruction routes of CH$_3$OH were introduced to the explosion model but it was found reactions during the explosion were not efficient enough to have a great effect.\par
The low efficiency of the reactions during the explosions, short explosion timescale and small abundance of parent species combined to give low abundances of the other COMs in the model. In the case of CH$_3$CHCH$_2$, the reaction rates are experimentally measured and so this failing of the model represents a major flaw. HCOOCH$_3$ was the most abundant of the under-produced COMs, though it formed in the post explosion gas phase from species released by the explosions.\par
Overall, this work shows that, based on our current understanding of the chemical network, it is unlikely that ice mantle explosions contribute significantly to the chemical composition of dark molecular clouds. The explosion model produces simple species equally well to a standard UCLCHEM model but underproduces most COMs and overproduces CH$_3$OH and CH$_3$CHO.\par
This poses a challenge as the models that included surface reactions but no explosions were similarly unable to match the observations. One solution may be found through laboratory measurements. The models of both processes have a large number of uncertain parameters and an improved agreement between the models and observations may be obtained if these are constrained. Alternatively, another formation process may be invoked for COM formation in cold gas. For example, the collision of dust grains in turbulent gas may lead to the synthesis of complex species \citep{Cassone2018} or cosmic rays may produce suprathermal molecules in ice mantles that can overcome reaction energy barriers to produce complex species \citep{Shingledecker2018OnModels}. \par
\acknowledgments
The authors thank the referees for their constructive comments which improved this manuscript. JH, SV and JMCR acknowledge funding from the STFC grant ST/M001334/1. NB and DS thank STFC for financially supporting their visit to UCL in July 2016.

\appendix
\section{Explosion Network}
The explosion is a high density, high temperature regime that is unusual for astrochemistry. The chemical network used in this work includes 143 radical-radical reactions that are believed to be possible. One key assumption is that, in the explosion conditions, the ``high-pressure limit'' applies. The underlying physical assumption is that the density is sufficiently high that a third body is always available to stabilise a product of a radical-radical reaction and so the reaction proceeds at the two-body rate.\par
Where possible, reaction routes and the relevant rates coefficients have been taken from the combustion chemistry literature. Formation routes were found for all of the COMs which have been observed in TMC-1 and are modelled in this work, except for HCOOCH$_3$. To complete the network, further reactions were assumed to be possible. Reactions involving two radicals are recombination reactions and only the products of their association are considered. In the case of the reactions between a radical and a closed shell molecule, the addition intermediate is assumed to be stabilized by the collision with the third body. Given that these assumed reactions have no known rate, they were given a random rate each time the model was run and many model runs were used to evaluate the resulting uncertainty in the abundances. The random rates are logarithmically sampled from the range \num{e-15} to \SI{e-9}{\centi\metre\cubed\per\second}. The reactions and any published rates that have been found are listed in Tables~\ref{table:explosionreactions1} to ~\ref{table:explosionreactions4}.
\begin{table*}
\caption{The explosion reaction network. Where rate coeffecients were available in the literature, a rate coefficient was calculated at \SI{1000}{\kelvin}. All other reactions used randomly sampled rate coefficents as noted in Section~\ref{sec:explosion-model}.}
\centering\begin{tabular}{llllll}
\hline
Reactants & & Products& & Rate / cm$^3s^{-1}$\\ 
\hline
OH & OH & H$_2$O$_2$ &  &\num{1.44e-11} &\citep{Sangwan2012}\\ 
OH & CH$_3$ & CH$_3$OH &  &\num{1.7e-10} &\citep{Jasper2007}\\ 
OH & CH$_2$ & H$_2$CO & H &\num{1.2e-10} &\citep{Jasper2007}\\ 
OH & CH & H$_2$CO &  &\num{e-15} - \num{e-9} &\\ 
OH & NH$_2$ & NH$_2$OH &  &\num{7.8e-11} &\citep{Klippenstein2009}\\ 
OH & NH & HNO & H &\num{4.1e-11} &\citep{Klippenstein2009}\\ 
OH & HCO & CO &  &\num{e-15} - \num{e-9} &\\ 
OH & H$_3$CO & CH$_3$OOH &  &\num{e-15} - \num{e-9} &\\ 
OH & CH$_2$OH & H$_2$CO & H$_2$O &\num{e-15} - \num{e-9} &\\ 
OH & CH$_2$OH & H$_2$CO & H$_2$O &\num{e-15} - \num{e-9} &\\ 
OH & NHOH & HNO & H$_2$O &\num{1.8e-11} &\citep{Sun2001}\\ 
OH & C$_2$H5 & CH$_3$CH$_2$OH &  &\num{1.3e-10} &\citep{Fagerstrom1993}\\ 
OH & CH$_3$NH & CH$_3$NHOH &  &\num{e-15} - \num{e-9} &\\ 
OH & C$_2$H$_3$ & CH$_2$CHOH &  &\num{e-15} - \num{e-9} &\\ 
OH & CH$_2$NH$_2$ & CH$_2$NH$_2$OH &  &\num{e-15} - \num{e-9} &\\ 
OH & CH$_2$CHO & CH$_2$OHCHO &  &\num{7e-11} &\citep{Tsang1986}\\ 
OH & CH$_3$OCH$_2$ & CH$_3$OCH$_2$OH &  &\num{e-15} - \num{e-9} &\\ 
OH & CH$_2$CH$_2$OH & (CH$_2$OH)$_2$ &  &\num{e-15} - \num{e-9} &\\ 
OH & CHNH & CHOHNH &  &\num{e-15} - \num{e-9} &\\ 
OH & N$_2$H$_3$ & N$_2$H$_3$OH &  &\num{e-15} - \num{e-9} &\\ 
OH & NHCHO & NHOHCHO &  &\num{e-15} - \num{e-9} &\\ 
OH & CH$_3$ONH & CH$_3$ONHOH &  &\num{e-15} - \num{e-9} &\\ 
OH & CH$_2$OHNH & CH$_2$OHNHOH &  &\num{e-15} - \num{e-9} &\\ 
CH & CH & C & CH$_2$ &\num{0.2e-10} &\citep{Bergeat1999}\\ 
CH & CH & C$_2$H & H &\num{1.8e-10} &\citep{Bergeat1999}\\ 
CH & NH$_2$ & CHNH$_2$ &  &\num{e-15} - \num{e-9} &\\ 
CH & NH & CHNH &  &\num{e-15} - \num{e-9} &\\ 
CH & HCO & CHCHO &  &\num{e-15} - \num{e-9} &\\ 
CH & H$_3$CO & CH$_3$OCH &  &\num{e-15} - \num{e-9} &\\ 
CH & C$_2$H6 &  &  &\num{e-15} - \num{e-9} &\\ 
CH & CH$_2$OH & CHCH$_2$OH &  &\num{e-15} - \num{e-9} &\\ 
CH$_2$ & CH$_2$ & C$_2$H$_2$ & H &\num{1.5e-10} &\citep{Jasper2007}\\ 
CH$_2$ & CH & C$_2$H$_3$ &  &\num{e-15} - \num{e-9} &\\ 
CH$_2$ & NH$_2$ & CH$_2$NH$_2$ &  &\num{e-15} - \num{e-9} &\\ 
CH$_2$ & NH & CH$_2$NH &  &\num{e-15} - \num{e-9} &\\ 
CH$_2$ & HCO & CO & CH$_3$ &\num{e-15} - \num{e-9} &\\ 
CH$_2$ & H$_3$CO & H$_2$CO & CH$_3$ &\num{e-15} - \num{e-9} &\\ 
CH$_2$ & CH$_2$OH & C$_2$H$_4$ & OH &\num{e-15} - \num{e-9} &\\ 
CH$_2$ & CH$_2$OH & H$_2$CO & CH$_3$ &\num{e-15} - \num{e-9} &\\ 
CH$_2$ & CHNH$_2$ & CH$_2$CHNH$_2$ &  &\num{e-15} - \num{e-9} &\\ 
CH$_2$ & CH$_3$OCH & CH$_3$OCHCH$_2$ &  &\num{e-15} - \num{e-9} &\\ 
\hline
\end{tabular}
\label{table:explosionreactions1}
\end{table*}
\begin{table*}
\caption{continued}
\centering\begin{tabular}{llllll}
\hline
Reactants & & Products& & Rate / cm$^3s^{-1}$ \\ 
\hline
CH$_2$ & CHCH$_2$OH & CH$_2$CHCH$_2$OH &  &\num{e-15} - \num{e-9} &\\ 
CH$_2$ & CHCHO & CH$_2$CHCHO &  &\num{e-15} - \num{e-9} &\\ 
CH$_2$ & C$_2$H$_4$ & CH$_3$CHCH$_2$ &  &\num{3e-14} &\citep{Laufer1975}\\ 
CH$_3$ & CH$_3$ & C$_2$H6 &  &\num{3.5e-11} &\citep{Wang2003}\\ 
CH$_3$ & CH$_2$ & C$_2$H5 &  &\num{2e-10} &\citep{Ge2010}\\ 
CH$_3$ & CH$_2$ & C$_2$H$_4$ & H &\num{2e-10} &\citep{Ge2010}\\ 
CH$_3$ & CH & C$_2$H$_4$ &  &\num{e-15} - \num{e-9} &\\ 
CH$_3$ & NH$_2$ & CH$_3$NH$_2$ &  &\num{1.5e-10} &\citep{Jodkowski1995}\\ 
CH$_3$ & NH & CH$_2$NH & H &\num{e-15} - \num{e-9} &\\ 
CH$_3$ & HCO & CH$_3$CHO &  &\num{5e-11} &\citep{Callear1990}\\ 
CH$_3$ & H$_3$CO & CH$_3$OCH$_3$ &  &\num{3e-10} &\citep{Balucani2015}\\ 
CH$_3$ & CH$_2$OH & C$_2$H5OH &  &\num{7.5e-11} &\citep{Sivaramakrishnan2010}\\ 
CH$_3$ & O & H$_3$CO &  &\num{1.4e-10} &\citep{Harding2005}\\ 
CH$_3$ & CH$_2$OH & CH$_3$CH$_2$OH &  &\num{e-15} - \num{e-9} &\\ 
CH$_3$ & NHOH & CH$_3$NHOH &  &\num{e-15} - \num{e-9} &\\ 
CH$_3$ & C$_2$H5 & CH$_3$CH$_2$CH$_3$ &  &\num{2.0e-11} &\citep{Mousavipour2003}\\ 
CH$_3$ & CH$_3$NH & CH$_3$NHCH$_3$ &  &\num{e-15} - \num{e-9} &\\ 
CH$_3$ & C$_2$H$_3$ & CH$_3$CHCH$_2$ &  &\num{e-15} - \num{e-9} &\\ 
CH$_3$ & CH$_2$NH$_2$ & CH$_3$CH$_2$NH$_2$ &  &\num{e-15} - \num{e-9} &\\ 
CH$_3$ & CH$_2$CHO & CH$_3$CH$_2$CHO &  &\num{e-15} - \num{e-9} &\\ 
CH$_3$ & CH$_3$OCH$_2$ & CH$_3$CH$_2$OCH$_3$ &  &\num{e-15} - \num{e-9} &\\ 
CH$_3$ & CH$_2$CH$_2$OH & CH$_3$CH$_2$CH$_2$OH &  &\num{e-15} - \num{e-9} &\\ 
CH$_3$ & CHNH & CH$_3$CHNH &  &\num{e-15} - \num{e-9} &\\ 
CH$_3$ & N$_2$H$_3$ & CH$_3$N$_2$H$_3$ &  &\num{e-15} - \num{e-9} &\\ 
CH$_3$ & NHCHO & CH$_3$NHCHO &  &\num{e-15} - \num{e-9} &\\ 
CH$_3$ & CH$_3$ONH & CH$_3$ONHCH$_3$ &  &\num{e-15} - \num{e-9} &\\ 
CH$_3$ & CH$_2$OHNH & CH$_3$NHCH$_2$OH &  &\num{e-15} - \num{e-9} &\\ 
HCO & NHOH & CHONHOH &  &\num{5.3e-11} &\citep{Xu2004}\\ 
HCO & C$_2$H5 & CH$_3$CH$_2$CHO &  &\num{3e-11} &\citep{Tsang1986}\\ 
HCO & CH$_3$NH & CH$_3$NHCHO &  &\num{e-15} - \num{e-9} &\\ 
HCO & C$_2$H$_3$ & CH$_2$CHCHO &  &\num{3e-11} &\citep{Tsang1986}\\ 
HCO & CH$_2$NH$_2$ & CH$_2$NH$_2$CHO &  &\num{e-15} - \num{e-9} &\\ 
HCO & CH$_2$CHO & CH$_2$(CHO)$_2$ &  &\num{e-15} - \num{e-9} &\\ 
HCO & CH$_3$OCH$_2$ & CH$_3$OCH$_2$CHO &  &\num{e-15} - \num{e-9} &\\ 
HCO & CH$_2$CH$_2$OH & CH$_2$CH$_2$OHCHO &  &\num{e-15} - \num{e-9} &\\ 
HCO & CHNH & CHCHONH &  &\num{e-15} - \num{e-9} &\\ 
HCO & N$_2$H$_3$ & N$_2$H$_3$CHO &  &\num{e-15} - \num{e-9} &\\ 
HCO & NHCHO & NH(CHO)$_2$ &  &\num{e-15} - \num{e-9} &\\ 
HCO & CH$_3$ONH & CH$_3$ONHCHO &  &\num{e-15} - \num{e-9} &\\ 
HCO & HCO & (CHO)$_2$ &  &\num{e-15} - \num{e-9} &\\ 
HCO & CH$_2$OH & CH$_2$OHCHO &  &\num{e-15} - \num{e-9} &\\ 
\hline
\end{tabular}
\label{table:explosionreactions2}
\end{table*}
\begin{table*}
\caption{continued}
\centering\begin{tabular}{llllll}
\hline
Reactants & & Products& & Rate / cm$^3s^{-1}$ \\ 
\hline
HCO & CH$_2$OHNH & CH$_2$OHNHCHO &  &\num{e-15} - \num{e-9} &\\ 
H$_2$CO & HCOH & HCOOCH$_3$ &  &\num{e-15} - \num{e-9} &\\ 
H$_3$CO & H &  & CH$_3$OH &\num{} &\citep{Xu2007}\\ 
H$_3$CO & CH$_2$OH & CH$_3$OCH$_2$OH &  &\num{e-15} - \num{e-9} &\\ 
H$_3$CO & NHOH & CH$_3$ONHOH &  &\num{e-15} - \num{e-9} &\\ 
H$_3$CO & C$_2$H5 & CH$_3$CH$_2$CH$_3$O &  &\num{e-15} - \num{e-9} &\\ 
H$_3$CO & CH$_3$NH & CH$_3$OCH$_3$NH &  &\num{e-15} - \num{e-9} &\\ 
H$_3$CO & C$_2$H$_3$ & CH$_2$CHOCH$_3$ &  &\num{e-15} - \num{e-9} &\\ 
H$_3$CO & CH$_2$NH$_2$ & CH$_3$OCH$_2$NH$_2$ &  &\num{e-15} - \num{e-9} &\\ 
H$_3$CO & CH$_2$CHO & CH$_3$OCH$_2$CHO &  &\num{e-15} - \num{e-9} &\\ 
H$_3$CO & CH$_3$OCH$_2$ & CH$_2$(CH$_3$O)$_2$ &  &\num{e-15} - \num{e-9} &\\ 
H$_3$CO & CH$_2$CH$_2$OH & CH$_3$OCH$_2$CH$_2$OH &  &\num{e-15} - \num{e-9} &\\ 
H$_3$CO & CHNH & CH$_3$OCHNH &  &\num{e-15} - \num{e-9} &\\ 
H$_3$CO & N$_2$H$_3$ & CH$_3$ON$_2$H$_3$ &  &\num{e-15} - \num{e-9} &\\ 
H$_3$CO & NHCHO & CH$_3$ONHCHO &  &\num{e-15} - \num{e-9} &\\ 
H$_3$CO & CH$_3$ONH & CH$_3$ONHCH$_3$O &  &\num{e-15} - \num{e-9} &\\ 
H$_3$CO & CH$_2$OHNH & CH$_3$OCH$_2$OHNH &  &\num{e-15} - \num{e-9} &\\ 
H$_3$CO & H$_3$CO & CH$_3$OCH$_3$O &  &\num{e-15} - \num{e-9} &\\ 
CH$_2$OH & CH$_2$OH & (CH$_2$OH)$_2$ &  &\num{e-15} - \num{e-9} &\\ 
CH$_2$OH & NHOH & CH$_2$OHNHOH &  &\num{e-15} - \num{e-9} &\\ 
CH$_2$OH & C$_2$H5 & CH$_3$CH$_2$CH$_2$OH &  &\num{2.0e-11} &\citep{Tsang1987}\\ 
CH$_2$OH & CH$_3$NH & CH$_3$NHCH$_2$OH &  &\num{e-15} - \num{e-9} &\\ 
CH$_2$OH & C$_2$H$_3$ & CH$_2$CHCH$_2$OH &  &\num{2.0e-11} &\citep{Tsang1987}\\ 
CH$_2$OH & CH$_2$NH$_2$ & CH$_2$NH$_2$CH$_2$OH &  &\num{e-15} - \num{e-9} &\\ 
CH$_2$OH & CH$_2$CHO & CH$_2$CHOCH$_2$OH &  &\num{e-15} - \num{e-9} &\\ 
CH$_2$OH & CH$_3$OCH$_2$ & CH$_3$OCH$_2$CH$_2$OH &  &\num{e-15} - \num{e-9} &\\ 
CH$_2$OH & CH$_2$CH$_2$OH & CH$_2$(CH$_2$OH)$_2$ &  &\num{e-15} - \num{e-9} &\\ 
CH$_2$OH & CHNH & CH$_2$OHCHNH &  &\num{e-15} - \num{e-9} &\\ 
CH$_2$OH & N$_2$H$_3$ & CH$_2$OHN$_2$H$_3$ &  &\num{e-15} - \num{e-9} &\\ 
CH$_2$OH & NHCHO & CH$_2$OHNHCHO &  &\num{e-15} - \num{e-9} &\\ 
CH$_2$OH & CH$_3$ONH & CH$_3$ONHCH$_2$OH &  &\num{e-15} - \num{e-9} &\\ 
CH$_2$OH & CH$_2$OHNH & (CH$_2$OH)$_2$NH &  &\num{e-15} - \num{e-9} &\\ 
CH$_3$OH & OH & CH$_2$OH & H$_2$O &\num{7.0e-12} &\citep{Li1996}\\ 
CH$_3$OH & OH & H$_3$CO & H$_2$O &\num{2.0e-12} &\citep{Li1996}\\ 
CH$_3$OH &  & HCOH & H$_2$ &\num{e-15} - \num{e-9} &\\ 
NH & NH & N$_2$H$_2$ &  &\num{1.0e-10} &\citep{Klippenstein2009}\\ 
NH & HCO & NHCHO &  &\num{e-15} - \num{e-9} &\\ 
NH & H$_3$CO & CH$_3$ONH &  &\num{e-15} - \num{e-9} &\\ 
NH & CH$_2$OH & CH$_2$OHNH &  &\num{e-15} - \num{e-9} &\\ 
NH & C$_2$H$_4$ & CH$_3$CHNH &  &\num{5.5e-12} &\citep{Mullen2005}\\ 
NH & CHNH$_2$ & CHNH$_2$NH &  &\num{e-15} - \num{e-9} &\\ 
\hline
\end{tabular}
\label{table:explosionreactions3}
\end{table*}
\begin{table*}
\caption{continued}
\centering\begin{tabular}{llllll}
\hline
Reactants & & Products& & Rate / cm$^3s^{-1}$ \\ 
\hline
NH & CH$_3$OCH & CH$_3$OCHNH &  &\num{e-15} - \num{e-9} &\\ 
NH & CHCH$_2$OH & CH$_2$OHCHNH &  &\num{e-15} - \num{e-9} &\\ 
NH & CHCHO & CHCHONH &  &\num{e-15} - \num{e-9} &\\ 
NH$_2$ & CH$_2$OH & NH$_2$CH$_2$OH &  &\num{e-15} - \num{e-9} &\\ 
NH$_2$ & NHOH & NH$_2$NHOH &  &\num{e-15} - \num{e-9} &\\ 
NH$_2$ & C$_2$H5 & CH$_3$CH$_2$NH$_2$ &  &\num{3.8e-11} &\citep{Lesclaux1978}\\ 
NH$_2$ & CH$_3$NH & CH$_3$NHNH$_2$ &  &\num{e-15} - \num{e-9} &\\ 
NH$_2$ & C$_2$H$_3$ & CH$_2$CHNH$_2$ &  &\num{e-15} - \num{e-9} &\\ 
NH$_2$ & CH$_2$NH$_2$ & CH$_2$(NH$_2$)$_2$ &  &\num{e-15} - \num{e-9} &\\ 
NH$_2$ & CH$_2$CHO & CH$_2$CHONH$_2$ &  &\num{e-15} - \num{e-9} &\\ 
NH$_2$ & CH$_3$OCH$_2$ & CH$_3$OCH$_2$NH$_2$ &  &\num{e-15} - \num{e-9} &\\ 
NH$_2$ & CH$_2$CH$_2$OH & CH$_2$CH$_2$OHNH$_2$ &  &\num{e-15} - \num{e-9} &\\ 
NH$_2$ & CHNH & CHNH$_2$NH &  &\num{e-15} - \num{e-9} &\\ 
NH$_2$ & N$_2$H$_3$ & NH$_2$NHNH$_2$ &  &\num{e-15} - \num{e-9} &\\ 
NH$_2$ & NHCHO & NH$_2$NHCHO &  &\num{e-15} - \num{e-9} &\\ 
NH$_2$ & CH$_3$ONH & CH$_3$ONHNH$_2$ &  &\num{e-15} - \num{e-9} &\\ 
NH$_2$ & CH$_2$OHNH & CH$_2$OHNHNH$_2$ &  &\num{e-15} - \num{e-9} &\\ 
NH$_2$ & NH$_2$ & N$_2$H$_4$ &  &\num{1.18e-10} &\citep{Klippenstein2009}\\ 
NH$_2$ & NH$_2$ & N$_2$H$_2$ & H$_2$ &\num{1.19e-19} &\citep{Klippenstein2009}\\ 
NH$_2$ & NH & N$_2$H$_3$ &  &\num{e-15} - \num{e-9} &\\ 
NH$_2$ & HCO & NH$_2$CHO &  &\num{e-15} - \num{e-9} &\\ 
NH$_2$ & H$_3$CO & CH$_3$ONH$_2$ &  &\num{e-15} - \num{e-9} &\\ 
NH$_2$ & CH$_2$OH & CH$_2$OHNH$_2$ &  &\num{e-15} - \num{e-9} &\\ 
\hline
\end{tabular}
\label{table:explosionreactions4}
\end{table*}
\bibliography{references.bib}
%
%
%
\end{document}